\documentclass[journal,twoside]{IEEEtran}
\IEEEoverridecommandlockouts
\usepackage{cite}
\usepackage{amsmath,amssymb,amsfonts}
\usepackage{algorithm,algorithmic,extarrows}
\usepackage{graphicx}
\usepackage{textcomp}
\usepackage{xcolor}
\usepackage{setspace}

\begin{document}
 \date{}
 \title{ Fundamental Tradeoffs in  Uplink Grant-Free  Multiple Access  with Protected CSI }
\author{
\IEEEauthorblockN{Dongyang Xu\IEEEauthorrefmark{1}\IEEEauthorrefmark{2}, Pinyi Ren\IEEEauthorrefmark{1}\IEEEauthorrefmark{2}, Yichen Wang\IEEEauthorrefmark{1}\IEEEauthorrefmark{2} and James A. Ritcey\IEEEauthorrefmark{3}}\\
 \IEEEauthorblockA{\IEEEauthorrefmark{1}School  of Information  and Communications Engineering, Xi'an Jiaotong University, China} \\
\IEEEauthorblockA{\IEEEauthorrefmark{2}Shaanxi Smart Networks and Ubiquitous Access Research Center, China. }\\
\IEEEauthorblockA{\IEEEauthorrefmark{3}Department of Electrical Engineering, University of Washington, USA.}\\
\IEEEauthorblockA{E-mail: \{\emph {xudongyang@stu.xjtu.edu.cn, \{pyren,wangyichen0819\}@mail.xjtu.edu.cn, ritcey@ee.washington.edu}
\}}
}
\maketitle
\begin{abstract}
In the envisioned 5G, uplink grant-free  multiple access will become the enabler of ultra-reliable low-latency communications (URLLC) services. By  removing the forward  scheduling request (SR) and backward scheduling grant (SG), pilot-based channel estimation and data transmission are launched in one-shot communications  with the aim of  maintaining the  reliability  of $99.999\% $ or more and latency of 1ms or less  under 5G new radio (NR) numerologies. The problem is that channel estimation can easily suffer from pilot aware attack which  significantly reduces  the system reliability. To solve this, we  proposed to apply the hierarchical 2-D feature coding (H2DF) coding  on time-frequency-code domain to safeguard channel state information (CSI), which  informs a fundamental rethinking of reliability, latency and accessibility. Considering uplink large-scale single-input multiple-output (SIMO)   reception of  short packets, we characterize the analytical closed-form  expression of reliability and define the accessibility of system. We find two  fundamental tradeoffs: reliability-latency and reliability-accessibility. With the the help of the two fundamental trade-offs, we demonstrate how  CSI protection could be integrated into uplink grant-free  multiple access to strengthen  URLLC services comprehensively.
\end{abstract}
\begin{IEEEkeywords}
URLLC,   grant-free  uplink  access, pilot-aware attack, reliability, accessibility.
\end{IEEEkeywords}
%
\newtheorem{lemma}{Lemma}
\newtheorem{property}{Property}
\newtheorem{definition}{Definition}
\newtheorem{proposition}{Proposition}
\newtheorem{principle}{Principle}
\newtheorem{assumption}{Assumption}
\newtheorem{problem}{Problem}
\newtheorem{theorem}{Theorem}
\newtheorem{remark}{Remark}
\newtheorem{fact}{Fact}
\newtheorem{question}{Question}
\newtheorem{example}{Example}
\section{Introduction}
\label{introduction}
With the  services and applications that require  stringent latency and reliability  emerging quickly, future fifth generation (5G) wireless networks  have defined a new category, i.e., low-latency and ultra-reliable  communication (URLLC),  to accommodate those services and applications comprehensively~\cite{ts22.261}.
To order to support URLLC services, air interface technologies are  experiencing an evolution of long term evolution (LTE)  and also a revolution towards 5G new radio (NR).  Among these, grant-free uplink   access is  recognized as  a key enabler  in which ultra-low latency is achieved by  novel orthogonal frequency
division multiplexing (OFDM) slot structures and  flexible/scalable numerologies while high reliability is maintained by advanced channel coding and  robust channel state estimation with margin for estimation error~\cite{Park}.

A preliminary  fact is  that conventional uplink transmission procedures consist of random access channel (RACH) operation, pilot-based channel estimation, scheduling request (SR) transmission, scheduling grant (SG) transmission and finally data transmission~\cite{Sesia}. By removing the procedures of SR and SG transmission, grant free uplink access  reduces the latency significantly and  also compels  channel estimation and data transmission to become more dependent on each other  for such  one-shot transmission. Take a blind detection scheme in grant-free Sparse code multiple access (SCMA)  in~\cite{Bayesteh}  for example in which   a  closed-loop SCMA system was  constituted by  active user detection, channel estimation and data decoding.  Pilot-based channel estimation provides  necessary feedbacks for performance  revisions and improvements  of iterative active user detection and  data decoding. Basically, the reliability of grant free uplink access severely relies on both  the resilience of pilot-based channel estimation and the reliability of data packet  decoding.
\begin{figure}[!t]
\centering \includegraphics[width=1.00\linewidth]{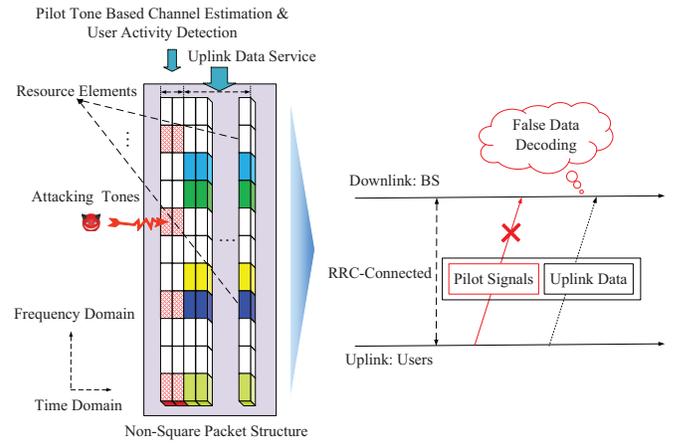}
\caption{Time-frequency domain illustration of pilot-aware attack on pilot-based channel estimation during uplink grant-free access.}
\vspace{-10pt}
\label{System_model_1}
\end{figure}

Pilot-based channel estimation requires that deterministic and publicly-known pilot tones  be shared on the time-frequency resource grid (TFRG) by all parties such that the receiver can recover the channels experienced by  pilot signals shared with the transmitter~\cite{Sesia}.  The first step to keep the resilience of channel estimation is to maintain the authenticity of pilots because applying the same pilot tones  at the receiver with the transmitter guarantees  the authenticity for channels estimated   by default. As shown in Fig.~\ref{System_model_1}, the critical problem  is that  a pilot-aware attacker which senses and acquires the public pilot information   can practically jamm/null/spoof those pilots  to destroy their authenticity, thus paralyzing the whole  regular channel estimation~\cite{Xu0}.   Due to the imprecise channel estimation,   the  data packet  decoding which requires accurate channel state information (CSI) as a prerequisite become beyond  being reliable.

Previously, it has been mentioned in ~\cite{Litchman} that  attacking  channel estimation is with high  priority  and more power-efficient than the behavior of jamming data packet decoding. Besides, the recent  accidents caused by  pilot aware attack indicate  that the conventional pilot-based channel estimation  is suffering  significant  risks~\cite{Xu1}. The commonsense built  to resolve this issue  is pilot randomization which transforms the pilot-aware attack into a  hybrid attack, including   \textbf{silence cheating (SC)},  \textbf{wide-band pilot jamming (WB-PJ)},  and \textbf{ partial-band pilot jamming  (PB-PJ)} attack. As a successful practice,  authors in~\cite{Xu1} proposed a  hierarchical 2-D feature coding (H2DF)-coding based pilot authentication  scheme  which can  protect pilot-based channel estimation  by  first delivering pilots  between transceiver pairs  in the form of diversified  subcarrier activation patterns (SAPs) and  then securing  its retrieval process. A valuable  result  is that  the time-domain consumption of  H2DF coding are  $K+1$ OFDM symbols provided  $K$  users simultaneously access the network.

 This give us a hint of applying H2DF coding  to safeguard  the  channel estimation of grant-free multiple access. Given  the constraint of  latency of 1ms or less,   this operation  however makes the   OFDM symbols  for  data transmission   change dynamically  with  the number of access users.  How to characterize the system   reliability and  is it possible to apply H2DF coding in grant-free multiple access? In what follows, we aim to answer  the above question and show  that our answer  is yes. The specific contributions  are summarized as follows:

 \begin{enumerate}
   \item  We propose a  novel  active user detection algorithm whereby perfect identification  of attack modes and  perfect detection of active users can be realized both.
   \item  We characterize a novel  expression of failure probability of uplink grant-free multiple access  in the regime of large-scale antenna arrays with a matched filter receiver. Based on this,  we find  a novel reliability-latency trade-off  under protected CSI.
   \item  We define the  $\xi$-accessibility of uplink grant-free multiple access with protected CSI as the ratio of the number of  active legitimate users that could maintain the failure probability less than $\xi$ to  the  total number of  OFDM symbols consumed during this period.  We  show that there exist a reliability-accessibility trade-off.
 \end{enumerate}

\section{ Basic Access Configuration and Attack Model}
\subsection{Packet and Frame Structure}
In URLLC, the latency that requires optimization consists of the processing latency $T_{\rm p}$ and the time to-transmit latency $T_{\rm t}$. $T_{\rm p}$ is constituted by three parts, i.e., the time of the channel acquisition, control channel decoding, and data detection. $T_{\rm t}$ is determined by the time-domain frame structure which depends on the non-square packet arrangement in the frequency domain. In order to minimize $T_{\rm p}$ and $T_{\rm t}$, a mini-slot structure proposed in [13] is considered, where each slot is divided into multiple mini-slots and each is equal to the transmission time interval (TTI) of $T_{\rm s}$. The corresponding frequency-domain subcarrier spacing is configured as ${\Delta f}$. $T_{s}$ and ${\Delta f}$ both follow the 5G NR numerologies.

Here, we consider the total latency ${T}$ constituted by the following:
\begin{equation}
{T} = \left( {m_e + m_d} \right){T_s}  + {T_{extra}}
\end{equation}
where $ m_e$ denotes the number of TTIs for channel estimation and  $ m_d$ denotes the number of  TTIs for data transmission.  ${T_{extra}}$ denotes  the time for other operations in BS and UE during the transmission.

\subsection{Uplink Grant-Free Multiple Access}
The basic uplink transmission procedures consist of random access channel (RACH) operation, pilot-based channel estimation, SR transmission,  SG transmission and finally data transmission. By removing the procedures of SR and SG transmission,  grant free  multiple access can reduce the latency significantly.

With this configuration, the RACH operation is responsible for establishing the connected state of radio resource control (RRC) between each transceiver pair. After that, each user prepares channel measurement and data transmission for the one-shot contention-based access. Since SG contains the scheduling decision for users, ignoring SG would compel BS to perform user activity detection before channel measurement to identify current user with access attempt. This functionality is realized by pilot signals for which exact pilot authentication and unique pilot allocation to each user guarantee the accuracy of user activity detection. For the data transmission,  the access strategy  includes but not limited to orthogonal frequency division multiple access (OFDMA). Different data streams from different users  share the same time-frequency resources.

To summary, the overall procedures include pilot-based active user detection, pilot-based channel estimation and OFDMA-based data transmission.
\begin{figure}[!t]
\centering \includegraphics[width=1.00\linewidth]{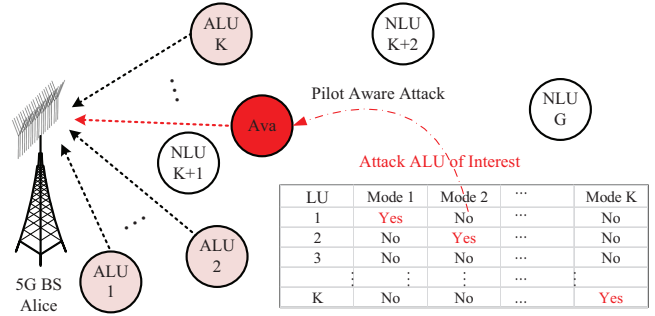}
\caption{Illustration of pilot-aware attack on uplink grant-free multiple access.}
\vspace{-10pt}
\label{System_model_2}
\end{figure}
\subsection{Pilot Randomization and  Pilot-Aware Attack}
Pilot randomization can avoid the pilot aware attack without imposing any prior information on the pilot design. The common method is to randomly select phase candidates. Each of the phase candidates is mapped by default into a unique quantized sample, chosen from the set $\cal A$ satisfying $\left\{ {{\phi}:{{{\phi} = 2m\pi } \mathord{\left/
 {\vphantom {{{\phi _k} = 2m\pi } C}} \right.
 \kern-\nulldelimiterspace} C},0 \le m \le C - 1}, C=\left| {\cal A} \right| \right\}$.

Under the circumstance of pilot randomization, pilot-aware  attacker can launch three arbitrary types of attacks, including WB-PJ attack,  PB-PJ  attack and SC.
The influence of pilot aware attack could be found in latency and reliability of grant-free multiple-access systems. Previous research ignored the influence of security risks on latency control and reliability preservation. In fact,   pilot-aware attack could easily paralyze   the authenticity of pilots and estimated channels. Without proper protection, the extra abnormal latency induced by the attack  would be extremely high and the reliability of  data packet decoding  based on the estimated channels would be also very low.



\section{System Model}
\label{SSPF}
Let us consider a  uplink single-input multiple-output (SIMO)-OFDM based grant-free multiple access system with a $N_{\rm T}$-antenna Alice and  $G$ single-antenna legitimate users (LUs) indexed by the set $\cal G$, $ \left| {\cal G} \right|= G$. As shown in Fig.~\ref{System_model_2}, LUs consist of $G-K$ non-active LUs (NLUs) and $K$ active LUs (ALUs) which are  indexed by the set $\cal K$ with $ \left| {\cal K} \right|= K$. Each of channels from ALUs to Alice  experience the  frequency-selective Rayleigh fading. The  pilot-tone based  channel estimation occupies    $m_e$ OFDM mini slots  and subcarriers within the set ${\Psi^{\rm{}}_{\rm{}}}$ satisfying  $\left| {\Psi _{\rm{}}^{}} \right|{\rm{ = }}{N_{\rm{}}}$   whereas the data transmission consumes  $m_d$ OFDM mini slots and subcarriers within the set ${\Psi _{\rm{D}}^{}}$ satisfying $\left| {\Psi _{\rm{D}}^{}} \right|{\rm{ = }}{N_{\rm{D}}}$. A single-antenna pilot-aware attacker  (named as Ava) could disturb pilot subcarriers  for channel estimation by launching a hybrid attack.

\begin{figure}[!t]
\centering \includegraphics[width=1.00\linewidth]{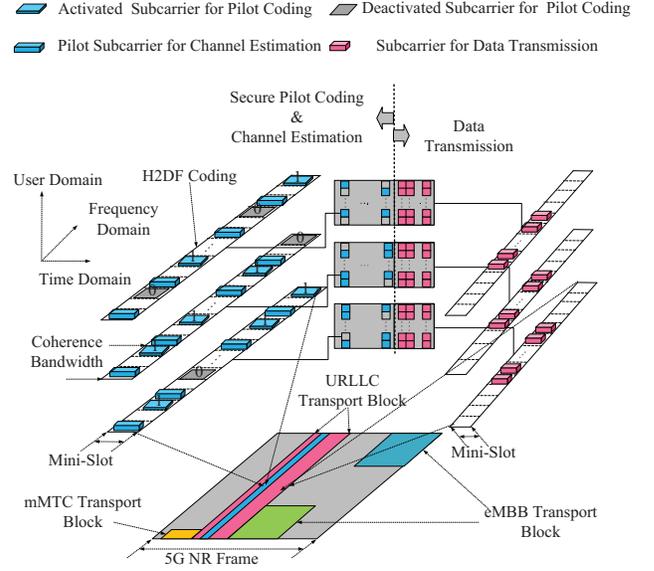}
\caption{Illustration of 5G NR frame structure with transport blocks respectively dedicated  for URLLC, mMTC and eMBB.   }
\vspace{-10pt}
\label{System_model_TVT}
\end{figure}

Consider H2DF coding based channel estimation to defend against pilot-aware attack in Fig.~\ref{System_model_TVT}.  Subcarriers within set ${\Psi^{\rm{}}_{\rm{}}}$ are dealt with separately.  One single  subcarrier for secure pilot coding   and one single pilot subcarrier  for channel estimation are respectively designated within each coherence bandwidth  but at different frequency-domain positions (i. e., two  pilot tones every three subcarriers). Therefore, we denote the set of  pilot subcarriers for  secure pilot coding  as ${\Psi^{\rm{}}_{\rm{R}}}$ satisfying  ${\Psi^{\rm{}}_{\rm{R}}}  \subseteq {\Psi^{\rm{}}_{\rm{}}}, \left| {\Psi _{\rm{R}}^{}} \right|{\rm{ = }}{N_{\rm{R}}}$ and the set of  pilot subcarriers for  channel estimation  as ${\Psi^{\rm{}}_{\rm{E}}}$ satisfying ${\Psi^{\rm{}}_{\rm{E}}}  \subseteq {\Psi^{\rm{}}_{\rm{}}},\left| {\Psi _{\rm{E}}^{}} \right|{\rm{ = }}{N_{\rm{E}}}$.
 \subsection{Pilot Signal Model for Channel Estimation}
We denote $x_{{\rm{L,}}m}^i\left[ k \right]$ and  $x_{\rm{A}}^i\left[ k \right]$ respectively as the  pilot tones  for the $m$-th ALU  and  Ava at the $i$-th subcarrier and $k$-th symbol time.   Pilot tones are  arranged following  a H2DF coding pattern  on ${\Psi^{\rm{}}_{\rm{R}}}$  whereas  those on ${\Psi^{\rm{}}_{\rm{E}}}$  follow the Block-type pilot insertion mode. On both ${\Psi^{\rm{}}_{\rm{R}}}$ and ${\Psi^{\rm{}}_{\rm{E}}}$,  the way of  inserting   pilot tones  across  subcarriers and OFDM symbols obeys the following principle, i.e.,   $x^{i}_{{\rm{L,}}m}\left[ k \right] = {x_{{\rm{L,}}m}}\left[ k \right] = \sqrt {{\rho _{{\rm{L,}}m}}} {e^{j{\phi _{k,m}}}}, {m \in {{\cal K}}}, \phi _{k,m} \in \cal A$ for either $ i \in {\Psi^{\rm{}}_{\rm{R}}}$ or $i \in {\Psi^{\rm{}}_{\rm{E}}}$.  $x^{}_{{\rm{L,}}m}\left[ k \right] $ can be  superimposed onto a dedicated  pilot sequence having been optimized under a non-security oriented scenario. The new pilot sequence can be utilized for channel estimation. At this point,  $\phi_{k,m}$ can be an additional phase difference for security consideration. We do not impose the phase constraint on the  strategies  of  pilot insertion at  Ava, that is,  $x^{i}_{{\rm{A}}}\left[ k \right] = \sqrt {{\rho _{\rm{A}}}} {e^{j{\varphi _{k,i}}}},i \in {\Psi^{\rm{}}_{\rm{}}}$.

Consider the basic OFDM procedure. We focus on the pilot tones on  subcarriers of  ${\Psi^{\rm{}}_{\rm{E}}}$.   Pilot tones of ALUs and Ava  over $N^{\rm }_{\rm E}$ subcarriers are  respectively  stacked as $N^{\rm }_{\rm E}$ by $1$  vectors ${{\bf{x}}_{{\rm{L,}}m}}\left[ k \right] = \left[ {{x^{j}_{{\rm{L,}}m}}\left[ k \right]} \right]_{j \in \Psi^{\rm }_{\rm E} }^{\rm{T}}$ and ${{\bf{x}}_{\rm{A}}}\left[ k \right] = \left[ {{x^{j}_{{\rm{A}}}}\left[ k \right]} \right]_{j \in \Psi^{\rm }_{\rm E} }^{\rm{T}}$.  The  length of cyclic prefix is assumed to be larger than  the maximum number  $L$ of channel taps.  The parallel streams, i.e.,  ${{\bf{x}}_{{\rm{L}},m}}\left[ k \right]$, $m \in \cal{K}$ and  ${{\bf{x}}_{{\rm{A}}}}\left[ k \right]$, are modulated with inverse fast Fourier transform (IFFT). Then the time-domain $N^{\rm }_{\rm E}$ by $1$  vector ${{\bf{y}}^i}\left[ k \right]$, derived by Alice after removing the cyclic prefix at the $i$-th receiving antenna, can be written as:
\begin{equation}\label{E.1}
{{\bf{y}}^i}\left[ k \right] = \sum\limits_{m = 1}^K {{\bf{H}}_{{\rm{C,}}m}^i{{\bf{F}}^{\rm{H}}}{{\bf{x}}_{{\rm{L,}}m}}\left[ k \right]}  + {\bf{H}}_{{\rm{C,A}}}^i{{\bf{F}}^{\rm{H}}}{{\bf{x}}_{\rm{A}}}\left[ k \right] + {{\bf{v}}^i}\left[ k \right]
\end{equation}
Here,  ${\bf{H}}_{{\rm{C}},m}^i$  is the $N^{\rm }_{\rm E} \times N^{\rm }_{\rm E}$ circulant matrices of the $m$-th ALU, with the  first column  given by ${\left[ {\begin{array}{*{20}{c}}
{{\bf{h}}_{{\rm{L,}}m}^{{i^{\rm{T}}}}}&{{{\bf{0}}_{1 \times \left( {N^{\rm }_{\rm E}- L_{\rm }} \right)}}}
\end{array}} \right]^{\rm{T}}}$.
${\bf{H}}_{{\rm{C,A}}}^i$ is a $N^{\rm}_{\rm E} \times N^{\rm }_{\rm E}$ circulant matrix with the first column given by ${\left[ {\begin{array}{*{20}{c}}
{{\bf{h}}_{\rm{A}}^{{i^{\rm{T}}}}}&{{{\bf{0}}_{1 \times \left( {N^{\rm A}_{\rm E} - L_{\rm s}} \right)}}}
\end{array}} \right]^{\rm{T}}}$. ${\bf{h}}_{\rm{L}, m}^i\in {{\mathbb C}^{L_{\rm } \times 1}} $ and ${\bf{h}}_{\rm{A}}^i \in {{\mathbb C}^{L_{\rm }\times 1}}$ respectively denote the   CIR  vectors  from  the $m$-th ALU and Ava to the $i$-th receive antenna of Alice. ${\bf{h}}_{\rm{A}}^i $ is  assumed to be independent with  ${\bf{h}}_{{\rm L},m}^i, \forall m \in \cal{K}$. ${\bf {F}}\in {{\mathbb C}^{N_{\rm E}^{\rm } \times N_{\rm E}^{\rm }}}$ denotes the discrete Fourier transform (DFT) matrix.   ${{\bf{v}}^i}\left[ k \right] \sim {\cal C N}\left( {0,{{{\bf{I}}_{N^{\rm }_{\rm E}}}\sigma ^2}} \right)$  denotes the noise vector on time domain at the $i$-th antenna of Alice within the $k$-th symbol time. $\sigma ^2$   is the average noise power of Alice.

Taking  FFT,  Alice  finally derives the    frequency-domain $N^{\rm }_{\rm E}$ by $1$ signal vector  at the $i$-th receive antenna as
\begin{equation}\label{E.3}
{\widetilde {{\bf{y}}}^i}\left[ k \right] =\sum\limits_{m = 1}^K {{{\bf{F}}_{\rm{L}}}{\bf{h}}_{{\rm{L}},m}^i{x_{{\rm{L}},m}}\left[ k \right]}  + {\rm{Diag}}\left\{ {{{\bf{x}}_{\rm{A}}}\left[ k \right]} \right\}{{\bf{F}}_{\rm{L}}}{\bf{h}}_{\rm{A}}^i + {{\bf{w}}^i}\left[ k \right]
\end{equation}
where  ${{\bf{w}}^i}\left[ k \right]={{\bf{F}}}{{\bf{v}}^i}\left[ k \right]$ and ${{\bf{F}}_{\rm{L}}} = \sqrt {N_{\rm E}^{\rm }} {\bf{F}}\left( {:,1:L_{\rm }} \right)$.
\subsection{Secure Pilot Coding Model}
 Let us briefly introduce the concept of H2DF coding. The H2DF-$\left( {K+1,B,C} \right)$ code of length $B$, size $C$ and order $K+1$ shown in~\cite{Xu1}, is denoted by  a $B\times C$ binary matrix  ${\bf{B}}{\rm{ = }}{\left[ {{{\bf b}_{j}}} \right]_{ {\rm{1}} \le j \le C}}$ where the $B\times 1$ vector ${\bf{b}}_{j}{\rm{ = }}{\left[ {{b_{i,j}}} \right]_{{\rm{1}} \le i \le B}}$  represents  its codeword.  Considering its definition~\cite{Xu1},  we know that: 1) every sum of up to $K+1$ different codewords can be decomposed by no codeword other than those used to form the sum; 2) every sum of up to $K+1$  different codewords is distinct from every other sum of $K+1$  or fewer codewords.
We spilt the columns  of code  matrix $\bf{B}$ into $K$ independent clusters. The $i$-th  cluster includes $\left[ {{C \mathord{\left/
 {\vphantom {C K}} \right.
 \kern-\nulldelimiterspace} K}} \right]$  columns indexed by  the set  ${\cal B}_{i}$ and  constitutes a sub-matrix denoted  by   $\left[ {{{\bf{b}}_{j \in {{\cal B}_{i}}}}} \right]$.   Each $i$-th sub-matrix  is  exclusively allocated  to the $i$-th ALU for mapping its randomized pilot phases to its SAPs in use.

 Basically, H2DF coding based pilot protection  is  a framework of random pilot phase encoding and decoding.   We define a pilot encoder as a map from $C$ discrete random phases into the  code  matrix $\bf{B}$.
 On this basis, the $i$-th ALU  independently conveys their own pilot phase  in the form of encoded SAPs which are programmed  by codewords ${{\bf{b}}_{i}}, i \in  {\cal K}$.  The specific principle is that   if  the $j$-th digit of the codeword  is equal to 1, the pilot tone signal  is inserted on the $j$-th subcarrier,  otherwise  this subcarrier will be  idle.  Those SAPs, after undergoing wireless channels, suffer from  the superposition interference from each other, and  finally  are superimposed and observed in the form of ${{\bf{b}}_{\rm{I}}}$  at Alice.  ${{\bf{b}}_{\rm{I}}}$ could be  separated and identified   by Alice  as   ${{\bf{b}}_{j}}, \forall j, 1\le j \le K+1$  which are recovered as  original pilots securely.

\section{User Activity Detection Under Hybrid Attack}
User activity detection  is the first step before channel estimation to determine  the set $\cal K$ of  ALUs that simultaneously perform uplink grant-free multiple access. This refers to the detection of the total number of ALUs and their  identities. Conventionally,  user activity detection  is  a blind detection process for which precise detection of  active user  rely on  multiple deterministic and distinguishable pilots. With those pilots being randomized,  user activity detection however comes to be a  main issue before channel estimation.

Under H2DF coding framework,  the user activity detection is rather a  different thing even though the channel estimation can be protected well. let's examine how to identify the total number of ALUs and their  identities under H2DF coding framework.   A fact is that H2DF coding creates a  unique temporal identifer for each LU such that Alice can identify  current  codewords in use  and their changes. This can be done by improving the order of H2DF coding to be $G+1$ and  defining the ${B \times \left( {\begin{array}{*{20}{c}}
C\\
k
\end{array}} \right)}$  matrix  ${{\bf{B}}_k}, k = 2, 3,..., G+1$, which is the collection of all of the Boolean sums (See Definition 1 in~\cite{Xu0}) of  codewords from $\bf B$, taken exactly $k$ at a time.   The $i$ column  of ${{\bf{B}}_k}$ is denoted by ${{\bf{b}}_{k,i}}$ and   represents a unique sum  codeword. As we can see,  every sum of up to $G+1$ different codewords can be decomposed by no codeword other than those used to form the sum and every sum of up to $G+1$  different codewords is distinct from every other sum of $G+1$  or fewer codewords. This means that each codeword  and each of its superposition versions with other codewords are both  unique. Therefore, the  identities of codewords  can be guaranteed. When  there is no attack, the total number of ALUs can also be detected precisely base on matrices ${{\bf{B}}_k}, k = 2, 3,..., G+1$.
\begin{algorithm}[!t]
\caption{User Activity Detection Under Hybrid Attack}
\begin{algorithmic}[1]
\STATE  According to the signal detection technique,  Alice  calculates  the number $N_j$ of pilot signals coexisting  at the $j, j\in {\Psi^{\rm{}}_{\rm{R}}}$-th subcarrier.
\STATE  Alice encodes the number  $N_j, j\in {\Psi^{\rm{}}_{\rm{R}}}$  as the binary codeword ${{\bf{b}}_{\rm{I}}}$.
\REQUIRE
\IF { all of the elements of ${{\bf{b}}_{\rm{I}}}$  are 1.  }
\STATE Indicate WB-PJ attack.   Encode the number  $N_j-1, j\in {\Psi^{\rm{}}_{\rm{R}}}$  as the new binary codeword ${{\overline{\bf{b}}}_{\rm{I}}}$ again.
\WHILE {$1\le j \le G+1$}
\STATE   Compare ${{\overline{\bf{b}}}_{\rm{I}}}$  with each column of matrix ${{\bf{B}}_{j}}$.
\IF {${{\bf{b}}_{\rm{I}}}$ belongs to the $k_0$-th column of  ${{\bf{B}}_{j_0}}$}
 \STATE   Break Down;
\ENDIF
\ENDWHILE
 \STATE  Indicate there exist $j_0$ ALUs and decompose ${{\bf{b}}_{k_0, j_0}}$ into the codewords of ALUs.
\ELSE
\WHILE {$1\le j \le G+1$}
\STATE   Compare ${{\bf{b}}_{\rm{I}}}$ with each column of matrix ${{\bf{B}}_{j}}$.
\IF {${{\bf{b}}_{\rm{I}}}$ belongs to the $k_1$-th column of  ${{\bf{B}}_{j_1}}$}
 \STATE   Break;
\ENDIF
\ENDWHILE
 \STATE  Output  ${{\bf{b}}_{k_1, j_1}}$. Indicate there exist $j_1$ ALUs.
 \IF {$\sum\limits_{i = 1}^{{N_{\rm{R}}}} {\sum\limits_{j = 1}^{{j_1}} {{b_{i,j}}} }  = \sum\limits_{j = 1}^{{N_{\rm{R}}}} {{N_j}}$}
 \STATE  Indicate SC and $j_1$ ALUs.  Decompose ${{\bf{b}}_{k_1, j_1}}$ into the codewords of ALUs.
 \ELSE
  \STATE  Find the set $\cal D$ in which for arbitrary $k \in {\cal D}$ each element $d_{k}$ makes $N_k=1$.
  \WHILE {$k \in {\cal D}$}
  \STATE Make $N_k=0$ and  encode the new number  $N_j, j\in {\Psi^{\rm{}}_{\rm{R}}}$  as the new binary codeword ${{\overline{\overline{\bf{b}}}}_{\rm{I}}}$.
  \WHILE {$1\le j \le G+1$}
\STATE   Compare ${{\overline{\overline{\bf{b}}}}_{\rm{I}}}$ with each column of matrix ${{\bf{B}}_{j}}$.
\IF {${{\overline{\overline{\bf{b}}}}_{\rm{I}}}$ belongs to the $k_2$-th column of  ${{\bf{B}}_{j_2}}$}
 \STATE   Break;
\ENDIF
\ENDWHILE
 \ENDWHILE
\STATE  Indicate PB-PJ attack and $j_2$ ALUs.  Decompose ${{\bf{b}}_{k_2, j_2}}$ into the codewords of ALUs.
 \ENDIF
\ENDIF
\ENSURE
\STATE The current attack mode; The number of ALUs; The codewords of ALUs.
\end{algorithmic}\label{Alogithm_PB}
\end{algorithm}

The signal detection technique is applied  on each subcarrier  to transform the signals into digits required for codewords. Consider arbitrary one (e.g., the $j$-th) pilot subcarrier. Firstly, Alice  stack the  receiving signals across $N_{\rm T}$ antennas within  $X+2,1\le X \le G$ OFDM symbols within the channel coherence time ($T_{c}$ OFDM symbols) and derive   $\left( {X+2} \right) \times {N_{\rm{T}}}$ receiving signal matrix by ${{\bf{Y}}_{\rm{D}}}$. Secondly, Alice derives  ordered eigenvalues of $\widehat {\bf{R}} = \frac{1}{{{\sigma ^2}}}{\bf{Y}}_{\rm D}{{\bf{Y}_{\rm D}}^{\rm{H}}}$  by  ${\lambda _{{X+2}}} >  \ldots  > {\lambda _1} > 0$ and constructs  the test statistics by $T = \frac{{{\lambda _{X+2}}}}{{{\lambda _{1}}}}\mathop {\gtrless}\limits_{{{{\overline{\cal H}}_0}}}^{{{{{ {\cal H}}_0}}}} \gamma$ where $\gamma$ is the  decision threshold.  The hypothesis ${{ {\cal H}}_0}$  means $X+2$ signals coexist and  ${{\overline{\cal H}}_{0}}$ means the opposite.  Authors in~\cite{Xu1}  provides  a function $\gamma \left( {{P_f}} \right) \buildrel \Delta \over = f\left( {{N_{\rm{T}}},G,{P_f}} \right)$ for determining on one subcarrier  the number of  antennas   required to achieve a  probability ${P_f}$ of  false alarm.  Alice could always expect a lower bound $\gamma \left( {{\varepsilon ^*}} \right)\le\gamma \left( {{P_f}} \right) $, $\gamma \left( {{\varepsilon ^*}} \right) = f\left( {{N_{\rm{T}}},G,{\varepsilon ^*}} \right)$  such that ${\varepsilon}^* \ge P_f$ is satisfied. Finally, Alice could select  $\gamma \left( 0 \right)$ as the detection threshold and  derive  new test statistics $T_{i} = \frac{{{\lambda _i}}}{{{\lambda _{1}}}}\mathop {\gtrless}\limits_{{{{\overline{\cal H}}_i}}}^{{{{{ {\cal H}}_i}}}} \gamma \left( 0 \right), 2\le i\le X+1$.  The hypothesis ${{ {\cal H}}_i}$  means $\left| {X +3- i} \right|$ signals coexist and  ${{\overline{\cal H}}_{i+1}}$ means the opposite. With that,  the number $N_{j}$ of coexisting signals on arbitrary one pilot subcarrier can be determined. For example,  two  signals  are recognized only  when  both ${{ {\cal H}}_{X+1}}$  and ${{\overline{\cal H}}_{X}}$ hold true.

Based on H2DF coding and signal detection technique, the next question is how  to perform user activity detection using ${{\bf{B}}_k}, k = 2, 3,..., G+1$ under the coexistence of WB-PJ attack, PB-PJ attack and SC. The perfect detection of user activity detection   is given in Algorithm~\ref{Alogithm_PB} for which energy detection and coding diversity should be jointly exploited.
\section{ Novel Characteristic  of Reliability Under Limited Block Lengths}
 Perfect active user detection  have been  a basis for channel estimation and data transmission which however  both operate with  a certain probability of failure.  The disturbance in reliability  of  channel estimation and  data transmission  becomes more severe  under the stringent requirements for latency, for example, $T_s=1 \rm ms$. Therefore, it is necessary to reevaluate the   reliability and  latency under H2DF coding framework.
\subsection{Reliability of Channel Estimation }
The reliability of channel estimation depends on  the reliability of pilot decoding securely.  With H2DF coding,  $B$  and $C$ will thus satisfy $ B = N_{\rm R}=q\left[ {1 +  {\left( {K+1} \right)}\left( {k - 1} \right)} \right],C = {q^k}, q \ge  {\left( {K+1} \right)}\left( {k - 1} \right)  \ge 3, K\ge 1$. $k$ is a predefined parameter which  is usually configured as 2 and 3. During  pilot decoding   process, the pilot identification error probability (IEP) represents its reliability.  Under maximum distance separable (MDS) code based  construction method,  the reliability of channel estimation, denoted by $P_{c}$, can be derived   as follows:
\begin{equation}\label{E.36}
P_{c}=\sqrt {\frac{{{{\left[ {1 + K\left( {k - 1} \right)} \right]}^k}}}{{2{{\left( {N_{{\rm R}}} \right)}^k}K}}}
\end{equation}
where
$N_{{\rm R}} \ge K\left( {k - 1} \right)\left[ {1 + K\left( {k - 1} \right)} \right]$,  $K\left( {k - 1} \right)\ge 3$.
\subsection{Decoding Error Probability of Short Packet Transmission}
Data decoding errors depend on the transmission rate $R$ in the occupied radio resources.  Without loss of generality, we assume:
\begin{enumerate}
  \item The attack on data transmission is not considered here  because pilot-aware attack is more preferred.
  \item The overloading factor of the system is not constrained, which means that both OFDMA and SCMA can be supported.  We also assume the same number of ALUs on each subcarrier for any  access strategy.
  \item All the ALUs  operate at the same rate $R$, calculated as $R = \frac{B}{{{N_{\rm{D}}}\Delta f \times {m_d}{T_{\rm s}}}}$.
  \item The  matched filter  is employed on each subcarrier.
\end{enumerate}
There are well-known expressions in the literature for calculating the outage probability for transmissions over fading channels as a function of the average signal-to-noise-ratio (SNR), the transmission rate $R$ and the receiver type~\cite{Berardinelli}. Consider  single-input-multiple-output (SIMO) Rayleigh fading on each subcarrier, the following relationship  between the normal approximation of the achievable rate $R$ and the decoding error probability  holds:
\begin{figure*}[!]
\normalsize
\begin{eqnarray}\setcounter{equation}{11}
\hspace{-10pt}
\frac{1}{{{N_{\rm{T}}}}}{\widehat {\bf{g}}}_{j,{m_0}}^H{{\bf{y}}_j}\left[ k \right] = \frac{{{d_{{\rm{L}},{m_0}}}\left[ k \right]}}{{{N_{\rm{T}}}}}\sum\limits_{i = 1}^{{N_{\rm{T}}}} {{{\left| {{{\widehat g}_{j,{m_0},i}}} \right|}^2}}  + \frac{{{d_{{\rm{L}},m}}\left[ k \right]}}{{{N_{\rm{T}}}}}\sum\limits_{m = 1,m \ne {m_0}}^K {\sum\limits_{i = 1}^{{N_{\rm{T}}}} {{\widehat g}_{j,{m_0},i}^ * } } {{\widehat g}_{j,m,i}} + \frac{1}{{{N_{\rm{T}}}}}\sum\limits_{i = 1}^{{N_{\rm{T}}}} {{\widehat g}_{j,{m_0},i}^ * } {w_{j,i}}\left[ k \right],1 \le {m_0} \le K
\end{eqnarray}
\vspace{-20pt}
\end{figure*}
\begin{figure*}[!]
\normalsize
\begin{eqnarray}\setcounter{equation}{15}
\hspace{-10pt}
{P_{e}} = {\left[ {1 - \left( {1 - \sqrt {\frac{1}{{2{q^k}K}}} } \right)Q\left( {\frac{{{\gamma _{asy}}\left[ {{{\log }_2}{\gamma _{asy}} - \frac{B}{{{N_{\rm{d}}}\Delta f \times \left( {T - m{T_{s}} - {T_{extra}}} \right)}}} \right]\sqrt {{N_d} \times \left( {T - m{T_{s}} - {T_{extra}}} \right)} }}{{\sqrt {{T_{s}}\left[ {\gamma _{asy}^2 - 1} \right]} }}} \right)} \right]^2}
\end{eqnarray}
\vspace{-20pt}
\end{figure*}
\begin{equation}\setcounter{equation}{5}
{P_d}= {\mathbb E}\left[ {Q\left( {\frac{{C\left( \gamma  \right) - R}}{{\sqrt {{{V\left( \gamma  \right)} \mathord{\left/
 {\vphantom {{V\left( \gamma  \right)} {\left( {{N_d}n} \right)}}} \right.
 \kern-\nulldelimiterspace} {\left( {{N_d}n} \right)}}} }}} \right)} \right]
\end{equation}
where $R = \frac{B}{{{N_{\rm{d}}}\Delta f \times n{T_{s}}}}$, $Q\left( x \right) = \int_x^\infty  {\frac{1}{{\sqrt {2\pi } }}} {e^{ - \frac{{{t^2}}}{2}}}dt$ and $C\left( \gamma  \right) = {\log _2}\left( {1 + \gamma } \right),V\left( \gamma  \right) = 1 - \frac{1}{{{{\left( {1 + \gamma } \right)}^2}}}$ with $\gamma $  defined as the instantaneous  SNR on each  subcarrier  in which there coexist $K_{c}$ interfering users. Without loss of generality,   we assume all the interferers are received with the same average $\gamma _0$ and the interference is Gaussian distributed.  The distribution ${f_{{K_c}}}\left( \gamma \right) $ of $\gamma $  under matched filter can be calculated by:
\begin{equation}
{f_{{K_c}}}\left( \gamma \right) = \frac{{{{\gamma}^{{N_T} - 1}}{e^{ - \frac{\gamma}{{{\gamma _0}}}}}}}{{\left( {{N_T} - 1} \right)!\gamma _0^{{K_c} + 1}}}\sum\limits_{i = 0}^{{N_T}} {\left( {\begin{array}{*{20}{c}}
{{N_T}}\\
i
\end{array}} \right)} \frac{{\gamma _0^{{K_c} + i}\Gamma \left( {{K_c} + i} \right)}}{{\Gamma \left( {{K_c}} \right){{\left( {\gamma+ 1} \right)}^{{K_c} + i}}}}
\end{equation}
where ${\Gamma \left( \cdot \right)}$ denotes the Gamma function. With this, the final decoding error probability can be calculated by:
\begin{equation}
{P_d} = \int_0^\infty  {{Q\left( {\frac{{C\left( \gamma  \right) - R}}{{\sqrt {{{V\left( \gamma  \right)} \mathord{\left/
 {\vphantom {{V\left( \gamma  \right)} {\left( {{N_d}n} \right)}}} \right.
 \kern-\nulldelimiterspace} {\left( {{N_d}n} \right)}}} }}} \right)}} {f_{{K_c}}}\left( \gamma \right)d{\gamma}
\end{equation}

\subsection{Reliability  for  Grant-Free Multiple Access}
To evaluate the  reliability of the whole  access process,   the reliability of  channel estimation and  decoding error probability  in data transmission should be both taken into consideration on the basis of  transmission strategies.

In this paper, at most one retransmission can be supported. The first transmission is  deemed  successful if  the  intended  ALU is correctly identified  and its data  is decoded successfully. In this case, the overall failure probability is $1- {\left( {1 - {P_{\rm{c}}}} \right){\left(1-P_{\rm{d}}\right)}}$.  When  the intended  ALU cannot be identified, or it is identified but its data can  not be  decoded, the ALU will perform a retransmission over shared resources. The probability of correctly decoding the retransmitted data can be  calculated by $\left[ {1 - {\left( {1 - {P_{\rm{c}}}} \right){\left(1-P_{\rm{d}}\right)}}} \right]{\left( {1 - {P_{\rm{c}}}} \right){\left(1-P_{\rm{d}}\right)}}$.
Finally the failure probability, denoted by ${P_{e}}$, is  given by:
\begin{equation}
{P_{e}} = {\left[ {1 - {\left( {1 - {P_{\rm{c}}}} \right){\left(1-P_{\rm{d}}\right)}}} \right]^2}
\end{equation}

\subsection{Asymptotic Results}
It is well understood that multi-antenna technique is instrumental to guarantee URLLC.  With  massive antennas equipped at BS,  the remarkable properties  tailored for URLLC   could be created, such as very high SNR links, quasi-deterministic links and extreme spatial multiplexing capability.

Consider the case presented here.  It is necessary to  evaluate the failure probability of  grant-free multiple access of short packets to massive-antenna  Alice  with protected CSI and under the constraint of ultra-low latency.  As the first step, we will give the basic receiving signal model at  each  subcarrier, for example, the $j$-th one, as follows:
\begin{equation}
{{\bf{y}}_j}\left[ k \right] = \sum\limits_{m = 1}^K {{{\bf{g}}_{j,m}}{d_{{\rm{L}},m}}\left[ k \right]}  + {{\bf{w}}_j}\left[ k \right]
\end{equation}
where  ${{d_{{\rm{L}},m}}\left[ k \right]},  m\in {\cal K}$ denotes  the symbol transmitted by the $m$-th ALU at the $k$-th OFDM mini-slot and satisfies $E\left[ {{{\left| {{d_{{\rm{L}},m}}\left[ k \right]} \right|}^2}} \right] = \gamma$. ${{\bf{w}}_j}\left[ k \right] = {\left[ {{w_{j,i}}\left[ k \right]} \right]_{1 \le i \le {N_{\rm{T}}}}},  j \in {{{\Psi} _{\rm{D}}}}$ denotes the noise vector at the $j$-th subcarrier and  the $k$-th OFDM mini-slot, and satisfies  ${{\bf{w}}_j}\left[ k \right] \sim {\cal C N}\left( {0,{{{\bf{I}}_{N_{\rm T}}}}} \right)$.
${{\bf{g}}_{j,m}}={\left[ {{g_{j,m,i}}} \right]_{1 \le i \le {N_{\rm{T}}}}} \in {{\mathbb C}^{{N_{\rm{T}}} \times 1}}, j \in {{{\Psi} _{\rm{D}}}}, m\in {\cal K}$ denotes the $j$-th subcarrier channel experienced by the $m$-th ALU and satisfies:
\begin{equation}
{{\bf{g}}_{j,m}} = {\left[ {\begin{array}{*{20}{c}}
{{{\bf{F}}_{{\rm{L,}}j}}{\bf{h}}_{{\rm{L}},m}^1}& \ldots &{{{\bf{F}}_{{\rm{L,}}j}}{\bf{h}}_{{\rm{L}},m}^{{N_{\rm{T}}}}}
\end{array}} \right]^T}
\end{equation}
where ${{{\bf{F}}_{{\rm{L,}}j}}}$ denotes the $j$-th row of ${{{\bf{F}}_{\rm{L}}}}$.

Two assumptions  are considered.  One is that precise channel estimation can be guaranteed and the other one is that estimation errors exist. By  defining ${\widehat {\bf{g}}_{j,m}} = {\left[ {{{\widehat g}_{j,m,i}}} \right]_{1 \le i \le {N_{\rm{T}}}}}$, receiving signals  after  matched filtering  are expressed  in Eq. (11).
Without loss of generality, take the $m_0$-th ALU for example. We have ${\widehat {\bf{g}}_{j,{m_0}}} = {{\bf{g}}_{j,{m_0}}}$ for the first case and  ${\widehat {\bf{g}}_{j,{m_0}}} = \left( {1 - \lambda } \right){{\bf{g}}_{j,{m_0}}} - \lambda {\widetilde {\bf{g}}_{j,{m_0}}},0 < \lambda  < 1$ for the second one where ${\widehat {\bf{g}}_{j,{m_0}}}\sim {\cal C N}\left( {0,{{{\bf{I}}_{N_{\rm T}}}}} \right)$ is independent with ${\widehat {\bf{g}}_{j,{m_0}}}$ and larger $\lambda $ means that channel estimation gets  worse.

Based on above preparations, we analyze  the asymptotic expression of $ \gamma$ as ${N_{\rm{T}}} \to \infty $.

\begin{theorem}
With precise channel estimated, the asymptotic expression of $ \gamma$ as ${N_{\rm{T}}} \to \infty $ is:
\begin{equation}\setcounter{equation}{12}
\gamma _{asy}\mathop  = \limits^\Delta \gamma _{asy}^{perfect}\xlongrightarrow[{N_{\rm{T}}} \to \infty]{ \rm{a.s.}} 1 + \frac{{{N_T}\gamma }}{{\gamma {{ K}_c} + 1}}
\end{equation}
and  the result  with  estimation errors  is:
\begin{equation}
\gamma _{asy}\mathop  = \limits^\Delta \gamma _{asy}^{error} \xlongrightarrow[{N_{\rm{T}}} \to \infty]{ \rm{a.s.}} 1 + \frac{{{N_T}\gamma \left( {{\rm{1 - }}\lambda } \right)}}{{\gamma {{ K}_c} + \lambda \gamma  + 1}}, {\rm{0 < }}\lambda  < {\rm{1}}
\end{equation}
Therefore, the decoding error probability  satisfies:
\begin{equation}
{P_d}\xlongrightarrow[{N_{\rm{T}}} \to \infty]{ \rm{a.s.}}{Q\left( {\frac{{C\left( \gamma _{asy}  \right) - R}}{{\sqrt {{{V\left( \gamma _{asy} \right)} \mathord{\left/
 {\vphantom {{V\left(\gamma _{asy} \right)} {\left( {{N_d}n} \right)}}} \right.
 \kern-\nulldelimiterspace} {\left( {{N_d}n} \right)}}} }}} \right)}
\end{equation}
The  failure probability of systems  can be finally expressed as the Eq.~(15). The specific value of $\gamma _{asy}$ depends on  the estimation assumption.
\end{theorem}

\subsection{Fundamental Tradeoffs}
Obviously, we can find a reliability-latency tradeoff  in Eq. (15). Besides this, we would like to give another tradeoff, that is, reliability-accessibility tradeoff.  Let us first define the accessibility which  is the practice of   keeping high reliability while being usable by as many  ALUs  as possible, operating within  specified tolerances.
\begin{definition}
We define the  $\xi$-accessibility of uplink grant-free multiple access with protected CSI as the ratio of the number of multiplexed ALUs   that could maintain $P_e$ less than $\xi$ to  the   number of  OFDM symbols for channel estimation and data decoding during this period,  that is,
\begin{equation}\setcounter{equation}{16}
{S} = \frac{K}{{{T-T_{extra}}}},  {P_{{e}}} \le \xi
\end{equation}
where $T-T_{extra}= \left( {K + 1 + {m_d}} \right){T_s} $ and  $\xi$ denotes the reliability constraint, usually equal to $10^{-5}$ when ultra-reliability is required.
\end{definition}

Obviously,  more accessibility will incur less reliability and vice versa.

\section{Numerical Results}
In this section, we evaluate the reliability performance of grant-free multiple access with protected CSI. We consider the frequency range 1 defined in~\cite{ts38.211} in which the channel bandwidth of 100MHz can be at most supported.  Subcarrier spacing is configured as 60kHz  and $T_{s}=17.86 \times 10^{-6}$ under 5G NR transmission numerologies defined in~\cite{ts38.211}. During the channel estimation,  two pilot tones every three subcarriers are configured with  one single subcarrier for secure pilot coding and one single pilot subcarrier for channel estimation, which is similar to the set-up in  LTE~\cite{Sesia}.  This  can support $N_{\rm R}=512=N_{\rm E}$ independent  subcarrier channels for H2DF coding with  $k=3$. For data transmission,  we consider $K$ ALUs  transmitting packets of  $R=32$ bytes over shared $N_{\rm D}=4$ subcarriers.   The number of channel taps is 6  and the  latency constraint  of $T \le 1 \rm ms$ is imposed.

\begin{figure}[!t]
\centering \includegraphics[width=1.00\linewidth]{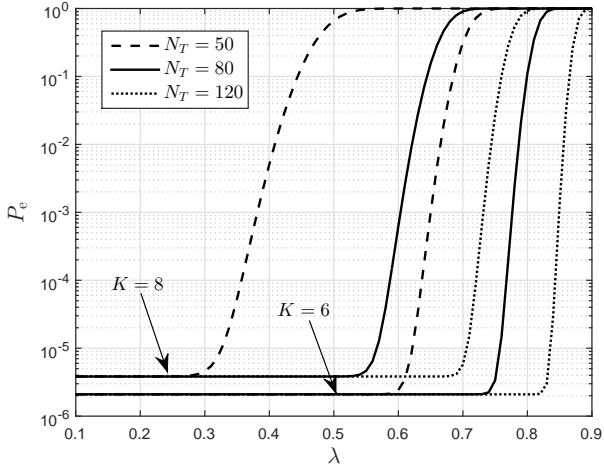}
\caption{The influence of channel estimation error  on the reliability of uplink grant-free multiple access  under different ALUs.}
\vspace{-10pt}
\label{Relibility_versus_error}
\end{figure}
\begin{figure}[!t]
\centering \includegraphics[width=1.00\linewidth]{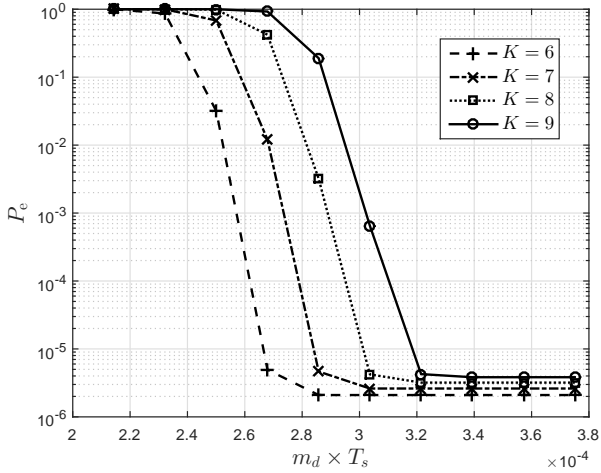}
\caption{Reliability-latency tradeoff under various number of ALUs.}
\vspace{-10pt}
\label{Relibility_versus_Latency}
\end{figure}
\begin{figure}[!t]
\centering \includegraphics[width=1.00\linewidth]{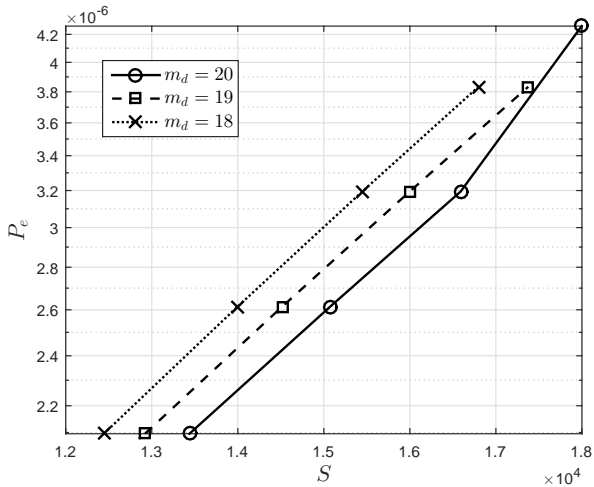}
\caption{Reliability-accessibility tradeoff under different number of mini-slots for data transmission.}
\vspace{-10pt}
\label{Relibility_versus_Accessibility}
\end{figure}

Fig.~\ref{Relibility_versus_error} presents the curve of $P_e$ versus the channel estimation error. The $ \gamma$ is fixed to be at 10 dB. As we can see, $P_e$ becomes increased with  increasing $\lambda$  above a certain threshold of error.  With the increase of number of antennas, this threshold of error comes to be less sensitive to its  changes.  The reason is that the  estimation error becomes less  with the increase of antennas.  The increase of ALUs also increases $P_e$. More AlUs introduce more uncertainty on channel estimation and more disturbance under estimation errors.

Fig.~\ref{Relibility_versus_Latency} shows the novel reliability-latency tradeoff curves under various number of ALUs. The number of antennas is fixed to be 100 and the $ \gamma$ is fixed to be at 20 dB. As we can see,  $P_e$  decreases  with the  increase of  latency  within a certain interval for guaranteeing  reliable data transmission. More  latency would not induce any improvement of reliability.

In Fig.~\ref{Relibility_versus_Accessibility}, we simulate the curves of reliability-accessibility tradeoff  under  different number of mini-slots for data transmission. The number of antennas is fixed to be 100 and the $ \gamma$ is fixed to be at 20 dB. To  achieve the reliability  less than  $10^{-5}$, the number of mini-slots for data transmission should be more than 18.  With those preparations,  we can see that the reliability  is decreased with the increase of accessibility.

\section{Conclusions}
\label{Conclusions}
In this paper, we proposed to apply H2DF coding to  conventional uplink grant-free multiple access to safeguard CSI. We showed that though  strengthening  reliability,  this operation could affect the  utilization of  time-frequency resources and therefore  we characterised a novel expression of reliability  and accessibility of this system. We could show that there exist a reliability-latency trade-off and a reliability-accessibility  trade-off. Those results presented  the possibility of satisfying  high reliability requirements for URLLC services while protecting CSI  and also gave  us hint of how to achieve this.

\end{document}